\documentclass[a4paper]{article}

\usepackage{icrc2013}
\usepackage[english]{babel}

\title{Very local interstellar spectra for galactic electrons, protons and helium}

\shorttitle{New very local interstellar spectra}

\authors{M.S. Potgieter
}

\afiliations{
Centre for Space Research, North-West University, Potchefstroom, South Africa\\ 
}

\email{Marius.Potgieter@nwu.ac.za}

\abstract{The local interstellar spectra (LIS) for cosmic rays at energies below $\sim$30 GeV/nuc are increasingly obscured from view at Earth by solar modulation, 
 the lower the energy. These charged particles encounter significant changes in the heliosphere, over an 11-year cycle, 
which include processes such as convection, diffusion, adiabatic energy losses and gradient, curvature and current sheet drifts. 
Particle drifts cause charge-sign dependent modulation and a 22-year cycle, adding complexity to determining the respective very LIS 
from observations only at Earth. However, with measurements now made by the Voyager 1 spacecraft in the vicinity of the heliopause, it is possible to determine a very 
LIS for galactic electrons between $\sim$5 MeV and $\sim$120 MeV. At these low energies also galactic protons observed in the outer heliosphere had been completely 
obscured by the so-call anomalous component which is accelerated inside the heliosheath. Since August 2012, these anomalous cosmic rays are 
substantially depleted at Voyager 1 so that also for cosmic ray ions it is now possible to obtain a lower limit to their very LIS. 
Combining numerical modelling of solar modulation with the accurate measurements by the PAMELA mission and with Voyager observations, 
the lower limit of the very LIS for electrons, protons and helium, and other ions, can be determined from $\sim$5 MeV and above. These spectra are called heliopause spectra, considered 
to be the lowest possible very LIS. Also from an astrophysics point of view, the determination of what can be called a very LIS, not just an averaged galactic spectrum, is encouraging. 
The mentioned aspects are discussed, focussing on a comparison of recent heliospheric observations and corresponding solar modulation modelling.}


\begin{document}
\maketitle

\section{Introduction}

Modelling and the subsequent understanding of the processes responsible for the solar modulation of cosmic rays (CRs) requires that an input spectrum must be specified at an assumed modulation boundary. 
This spectrum, usually called a very local interstellar spectrum (LIS), is then modulated throughout the heliosphere as a function of position, energy and time. 
Computed modulated spectra are usually compared to observations close to or at Earth (satellite and balloon experiments) and along trajectories of spacecraft such as Voyager 1 and 2, Ulysses, and others, 
to study the processes that are responsible of solar modulation, in fine detail.

The focus of these global approaches to solar modulation is to understand what exactly is causing solar modulation with emphasis on convection, diffusion, adiabatic energy losses and gradient, curvature and current sheet drifts. 
Apart from the 11-year cycle, particle drifts cause a 22-year cycle with charge-sign dependent modulation, adding complexity in determining the respective very LIS.
These processes are governed by a combination of modulation parameters including a full diffusion and drift tensor, the solar wind speed and its spatial dependence, which governs adiabatic energy changes,
the heliospheric magnetic field with its embedded turbulence and global geometry, 
and of course the geometry of the heliosphere and its main features such as the solar wind termination shock, the heliosheath and heliopause, and where the modulation boundary is actually located. 
For a comprehensive review on these aspects, see [1]. 

\begin{figure*}[!t]
\centering
\includegraphics[width=0.7\textwidth]{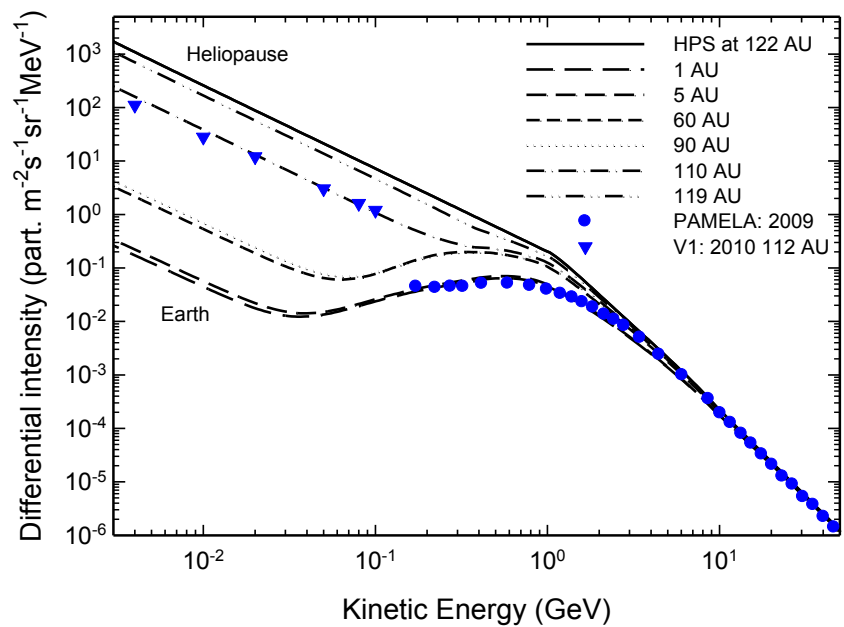}
			\caption{Computed modulated electron spectra at Earth (1 AU; with polar angle of $\theta = 90^{\circ}$), at 5 AU, 60 AU, 90 AU, 110 AU and 119 AU (with $\theta = 60^{\circ}$ 
			corresponding to the Voyager 1 trajectory). The modulation of electrons up to Earth was discussed in detail by [3]. The computed spectra are compatible with observations from Voyager 1 (2010) 
			and from PAMELA during the solar minimum of 2009, as indicated. A heliopause spectrum (HPS, upper solid line) is computed as required at the heliopause, positioned in the model at 122 AU. 
			A power law with $E^{-(1.5\pm0.1)}$ is found for these low energies with a clear spectral break between $\sim$800 MeV and $\sim$2 GeV to match $E^{-(3.15\pm 0.05)}$ at high energies.
			 Accordingly, the prediction is that the HPS at 10 MeV has a value of between 200 to 300 electrons m$^{-2}$s$^{-1}$sr$^{-1}$MeV$^{-1}$.}
\label{fig:Figure1}
\end{figure*}

Establishing the three major diffusion coefficients and particle drifts, as a function of time, space and energy, is a work in progress. Since these parameters are not uniquely 
and globally known, over a full 11-year cycle, the very LIS cannot be determined by using only observations at or close to Earth 
(despite the user friendliness of the force-field modulation approach based on limited physics). All attempts of computing cosmic ray LIS at kinetic energies $E < 10$ GeV from observations at Earth remain controversial. 
Progress is however made because solar modulation models and knowledge of  the underlying heliospheric turbulence have reached a relatively high level of sophistication.  	

Most helpful in this context is that CR observations are now becoming available from Voyager 1 and 2, both positioned deep inside the inner heliosheath, with Voyager 1 probably already beyond the heliopause [2],
and now entering the local interstellar medium. When these CR spectra as observed close to the heliopause and those close to and at Earth are compared to computed spectra over the full energy range relevant to 
solar modulation, (from 1 MeV to 50 GeV), the global modulation parameters can be determined much better so that trustworthy predictions about all the very LIS can be made.

These required input spectra, specified at the modulation boundary usually assumed to be the heliopause, should rather be called heliopause spectra (HPS), surely not galactic spectra (GS). 
In this context it should be mentioned that computed GS do not usually contain the contributions of any specific (local) sources within parsecs from the heliosphere 
so that an average GS may be different from an interstellar spectrum which may be different from a LIS (thousands of AU away from the Sun), which might be different from a very LIS or what can be called a HPS. 
The HPS are the CR spectra right at the edge of the heliosphere, say within $\sim$150 AU from the Sun in the direction the heliosphere is moving, obviously the ideal spectra to be used as input spectra for solar modulation models.

\begin{figure*}[!t]
\centering
\includegraphics[width=0.7\textwidth]{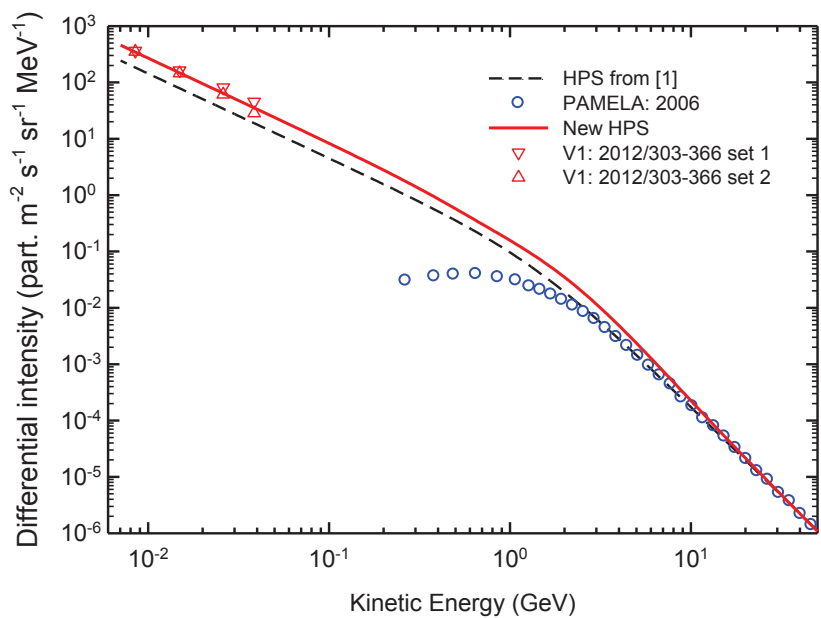}
			\caption{A comparison between the new electron HPS (solid red line) and a previously predicted HPS (dashed line) with a 
spectral index of $−(3.15\pm0.05)$ above 2 GeV and normalized to the PAMELA data above 30 GeV. 
The preliminary PAMELA electron intensities from November 2006 are presented by the blue circles. 
The HPS indicated by the dashed line, is the predicted HPS by [3] which did not take into account the 'bump' 
between $\sim$2 GeV and $\sim$20 GeV, also not the case in figure 1. The 'bump' requires a different spectral slope than the one above 30 GeV.
The red solid line is the new HPS in comparison with Voyager 1 electron data consisting of two sets of spectra, the first with 
an index of $-1.36$ and the second with $-1.53$ [see 13]. The unexpected ``bulge'' in the electron spectrum between $\sim$2 GeV and up to $\sim$20 GeV 
produces a spectral index of $-(3.3\pm 0.1)$ instead of $-(3.15\pm 0.05)$ in the mentioned energy range.
This means that the HPS or very LIS for electrons below $\sim$50 GeV may have more than just the one spectral break as was shown in figure 1.}
\label{fig:Figure2}
\end{figure*}

Attempting to make progress in this regard concerning galactic electrons, the solutions of comprehensive numerical models for the modulation of these electrons in the heliosphere were
presented in comparison with Voyager 1 and PAMELA observations by Potgieter and Nndanganeni [3]. For an attempt to describe the modulation parameters in the heliosheath, see also Nkosi et al. [4]. 
Their work contained observational data from Voyager 1 [5] and the PAMELA mission [6,7], with reference to the detailed electron spectra (but still preliminary) from PAMELA as reported by Di Felice [8], Munini [9] and Vos [10]. 
The final electron and positron spectra from PAMELA for 2006 to 2009 should become available soon. (See also the supporting and complimentary reports on electron observations and modelling made during the Rio de Janeiro ICRC conference).  
Potgieter et al. [11,12] gave updates during this conference of the mentioned work on an electron HPS, following up on previous publications made before the latest publications in Sciencexpress of the Voyager team [e.g.13,14]. 

In this short review, the updated HPS for electrons is shown and discussed, together with HPS for galactic protons and helium at these low energies, possible again because of Voyager 1 being at the heliopause. 
These new HPS are compared to computed galactic spectra mostly based on using GALPROP calculations [e.g. as described in 23].

\section{Modulation Models}

The purpose of this review is not to discuss numerical models for the solar modulation of cosmic rays so that only some basic information is given. 
It suffices to mention that a full three-dimensional (3D) numerical model, with three spatial dimensions and an energy dependence (four computational dimensions), 
had been used to compute electron, proton and helium spectra at selected positions in the heliosphere, including the inner heliosheath. 
The details of these model were published by [e.g. 1, 3,15,16]. 

For reviews on the prime geometrical and physical features of the heliosphere as applied to solar modulation, done with MHD and HD models, see e.g. Ferreira et al. [17],
Strauss et al. [18] and Luo et al. [19]. 

Transport models are based on the numerical solution of the heliospheric transport equation [20] for solar modulation including all four major modulation processes with a full diffusion and drift tensor, 
so that diffusion and drifts are not arbitrarily handled. The detail on the spatial and rigidity dependence of the three major diffusion coefficients and the drift coefficient is quite important, 
but also not reproduced here [see 3,15].

It is usually assumed that solar modulation becomes negligible with $E > 50$ GeV (see [21]). 

The heliopause is typically varied between 120 to 122 AU in these models with the TS at 94 AU, which gives a 26 to 28 AU wide heliosheath in the direction 
in which Voyager 1 has been moving. The TS is of course moving in-and-outwards with changing solar activity. This width is a determining factor in the modulation 
of galactic electrons between 1 MeV and 100 MeV [e.g. 4], and crucially important for the existence of the so-called anomalous component [1,27]. 

Lately, the focus has been on solar minimum modulation, with 2009 now confirmed as an exceptional good example of such an epoch [15].

\section{Galactic electrons}

First, the focus is on the HPS presented as very LIS for galactic electrons.The mentioned numerical model and the available Voyager 1 observations of galactic electrons 
between $\sim$6 MeV to $\sim$120 MeV during 2010 was used to determine a HPS which can be considered a very LIS below 100 MeV [16]. 
Until the recent Sciencexpress publications on Voyager 1 observations as mentioned above, the electron spectrum observed in 2010 was an optimal choice from the available Voyager 1 spectra 
because earlier observed spectra, closer to the TS, are much lower and may be subject to short-term changes as the region closer to the TS is more turbulent as it shifts position with changing solar activity. 
During this period Voyager 1 moved between 112 AU and 115 AU from the Sun. 

Using appropriate modulation parameters for the heliosheath and the mentioned 2010 spectrum, 
an input spectrum was determined between 1 MeV and 200 MeV and presented as a HPS. This was done by determining the spectral slope at these energies.
Then the differential flux values between 1 MeV and 200 MeV were computed to reproduce the 2010 Voyager spectrum, at the same time addressing the amount of modulation 
that takes place between the modulation boundary (typically the heliopause) and the observational positions. 

The spectral slope of this spectrum was subsequently adjusted with a different power law index above $\sim$1 GeV to reproduce the PAMELA spectra at Earth.
These spectra unfortunately extend down to only $\sim$200 MeV whereas the Voyager 1 measurements extend only up to $\sim$120 MeV. 
All the modulation parameters (including the solar wind speed and solar magnetic field) as required to reproduce the Voyager 1 and PAMELA observations are given and motivated by [3].

The modulated spectra with respect to the computed HPS are shown in figure 1. The computed electron spectra at Earth (1 AU, with polar angle of $\theta = 90^{\circ}$), at 5 AU, 60 AU, 90 AU, 
110 AU and 119 AU (with $\theta = 60^{\circ}$ corresponding to the Voyager 1 trajectory) are shown. 
The computed spectra at Earth and at 110 AU are compatible with corresponding observations from Voyager 1 and PAMELA (an average for 2009) as indicated. 
The HPS (upper solid line) is specified at the heliopause, positioned in the model at 122 AU, with the TS at 94 AU. These computed spectra illustrate how galactic electrons
are modulated, quite substantially in the heliosheath inwards to 90 AU, then significantly less to 60 AU. Inside 30 AU, low energy electrons are completely dominated by Jovian electrons, 
not shown in figure 1 (for an elaborate discussion, see [3]) so that the intensity of galactic electrons is not known at Earth below about 50 MeV.

Concerning the HPS, the following results were found: To reproduce the spectral shape of the Voyager 1 spectrum for 2010, the LIS below $\sim$1.0 GeV must have a power law form with $E^{-(1.55\pm0.05)}$, 
where $E$ is kinetic energy. To reproduce the PAMELA electron spectra observed between mid-2006 and the end of 2009, the HPS must have a power law form above $\sim$5 GeV with $E^{-(3.15\pm 0.05)}$. 
This HPS thus consists of two power laws with a 'break' occurring between $\sim$800 MeV and $\sim$2 GeV. This is consistent with the reported power law with a spectral index 
of $-(3.18\pm 0.05)$ by the PAMELA team [7]. The HPS reported here at low energies is different (significantly higher) from the interstellar spectra preferred by Webber and Higbie [22].

It should also be kept in mind that the Voyager 1 and 2 detectors cannot distinguish between electrons and positrons whereas the PAMELA detector can.
By the end of 2013, Voyager 1 will be 126.4 AU away from the Sun while Voyager 2 will be at 103.6 AU, both moving into the nose region of the heliosphere
but at a totally different heliolatitude (http://voyager.gsfc.nasa.gov/heliopause/data.html). Future Voyager 1 observations should enlighten us as it moves outwards at 3 AU per year.

For diffusion coefficients independent of energy for electrons at these low energies as predicted by turbulence theory and discussed by [3], this power law for the HPS is maintained as the spectra become modulated.
Introducing any form of energy (rigidity) dependence will not give steady modulation (difference between the modulated spectrum and the LIS is unchanged) over an energy range from 1 MeV to $\sim$ 200 MeV beyond 110 AU.
	
Evident from figure 1 is that the heliosheath acts as a very effective modulation 'hurdle' for these low energy electrons, causing intensity radial gradients of up to 20$\%$ AU$^{-1}$ as modelled by [4]. 
The bulk of the modulation for these electrons occurs between the heliopause and 80-90 AU, effectively inside the inner heliosheath. 
For an illustration of how the galactic intensity of 12 MeV may change with decreasing distance towards the Sun, see [3].

It appears from electron counting rates (not differential flux) reported by [5] that Voyager 1 observed a significant increase over a short distance what appears to be the heliopause. This large increase was quite unexpected.
During the ICRC newly observed galactic electron spectra were presented by the Voyager team [see also Ed Stone's highlight talk/paper] as published in Sciencexpress at the end of June 2013 [13,14]. 
They investigated a new method of analysing electron data to produce differential intensities (spectra) and reported that such systematic uncertainties influence 
the spectral shape so that the very clear -1.5 spectral index evident from the 2010 Voyager spectrum [3,16]
is to be modified to $E^{-(1.45\pm0.09)}$. The observed HPS presented by [13] are consistent with the prediction by [12] but higher than predicted by [3].

Concerning the break in the spectral shape of the presented HPS, Strong et al. [23] re-analysed synchrotron radiation data using various radio surveys to constrain the low-energy electron 
LIS in combination with data from Fermi-LAT and other experiments (see also [32] for earlier attempts). They concluded that the electron LIS exhibits a spectral break below a few GeV in accord with what is presented here. 
In addition they stated that the LIS below a few GeV has to be lower than what standard galactic propagation models predict, again consistent to what is shown here, and speculated about the causes of this result. 
It should be noted that the relation between synchrotron data and electrons is complicated by the presence of secondary electrons and positrons in the Galaxy 
so that it is not a straightforward matter to determine the spectral slope below a few GeV with galactic propagation modelling.
 
Further investigation from an astrophysics point of view seems required to establish if the $E^{-(1.50\pm0.15)}$ spectral slope below a few GeV as found here, has a galactic 
origin and as such presents a challenge to CR source models. For example, if the rigidity dependence of the diffusion coefficient in a model such as GALPROP would be adjusted,
could it explain this observed power law which is different from what is normally used in such models as the injection (source) spectrum for electrons?
Or, to speculate further from a heliospheric point of view, is this power-law perhaps influenced by the solar wind TS? It is rather curious that if the TS would be considered a strong, planar type shock with a compression ratio of 
$s$ = 4, the spectral shape of the consequent TS spectrum is $E^{-(1.0)}$, whereas for a weaker TS, with $s$ = 2.5, it becomes $E^{-(1.5)}$. 
The TS is much more complicated than a plain shock, so that this aspect, although unlikely, requires further study.

A comment on CR modulation beyond the HP is that it has become a very relevant topic since both Voyager spacecraft are about to explore the 
outer heliosheath (beyond the HP) and therefore may actually measure a pristine LIS sooner than what we anticipate. In this context Scherer at al. [24] argued that a 
certain percentage of CR modulation may occur beyond the HP. Recently, Strauss et al. [18] followed this up and computed that the differential intensity of 100 MeV protons may decrease by $\sim 25\%$
from where the heliosphere is turbulently disturbed (inwards from the heliospheric bow wave) up to the HP. However, because the diffusion coefficients 
of low energy electrons are independent of energy, making them significantly larger than for protons of the same rigidity, this percentage should be 
much less for 100 MeV electrons. 

Further study of the preliminary electron spectra from PAMELA for the period 2006 to 2009 as reported by [8,9,10] identified a new feature in all the observed electron spectra which is not evident from figure 1. 
These spectra exhibit a spectral feature (bump) between $\sim$2 GeV and $\sim$20 GeV as a complication to the nice (and simple) picture presented above as the electron LIS consisting of only two power laws.
See also [11]. 

This new feature is shown in figure 2 where a comparison is made between the PAMELA spectrum for November 2006 and a HPS with a spectral index of -3.15 above 2 GeV, 
normalized to the PAMELA data above 30 GeV but with a power law of $E^{-(1.5)}$ at low energies. According to Adriani et al. [7], who conducted a study of electron measurements made 
by PAMELA from July 2006 to January 2010, between $1\textnormal{ GeV}$ and $625\textnormal{ GeV}$, 
a spectral index of $-(3.18\pm 0.05)$ is required for electrons above 30 GeV. Shown in figure 2 is that when a spectral index of -3.15 is used above 2 GeV, as in figure 1, 
the corresponding HPS is clearly below the observed differential flux values observed at Earth between $\sim$2 GeV and $\sim$10 GeV.  
Of course, a HPS or a very LIS that is lower than the flux at Earth is unacceptable from a modulation point of view. 
At these energies, the modulation is already significant enough so that the observed values should always be below the very LIS. 
Clearly, the $E^{-3.15}$ dependence at higher energies is obscured by an unusual range of enhanced intensities observed between $\sim$2 GeV and up to $\sim$20 GeV. 
This means that the HPS or very LIS must be adjusted in this energy range to have a different spectral index than the reported -3.15 while keeping this index above $\sim$20 GeV.
This unexpected ``bulge'' in the electron spectrum between $\sim$2 GeV and up to $\sim$20 GeV produces a spectral index of $-(3.3\pm 0.1)$ instead of $-(3.15\pm 0.05)$ in the mentioned energy range.
The implication is that the HPS or very LIS for electrons below $\sim$50 GeV may have more than just the one, rather clear, spectral break, as was shown in figure 1.

The HPS shown as the solid (red) line in figure 2 takes this observed bump into account as well as the observed Voyager 1 differential intensity around 10 MeV, using a power law of  $E^{-(1.5)}$ at lower energies.
The dashed line is the prediction by [3]; see also [12].
The Voyager 1 electron data consist of two sets of spectra, the first with a power index of -1.36 and the second with -1.53 [see 13] based upon two different techniques to calculate the differential intensity.
This spectrum is presented as a HPS at a distance of 122 AU from the Sun, and as such the lowest possible very LIS. Any calculated galactic electron spectrum should therefore take this into account.

\begin{figure*}[!t]
\centering
\includegraphics[width=0.65\textwidth]{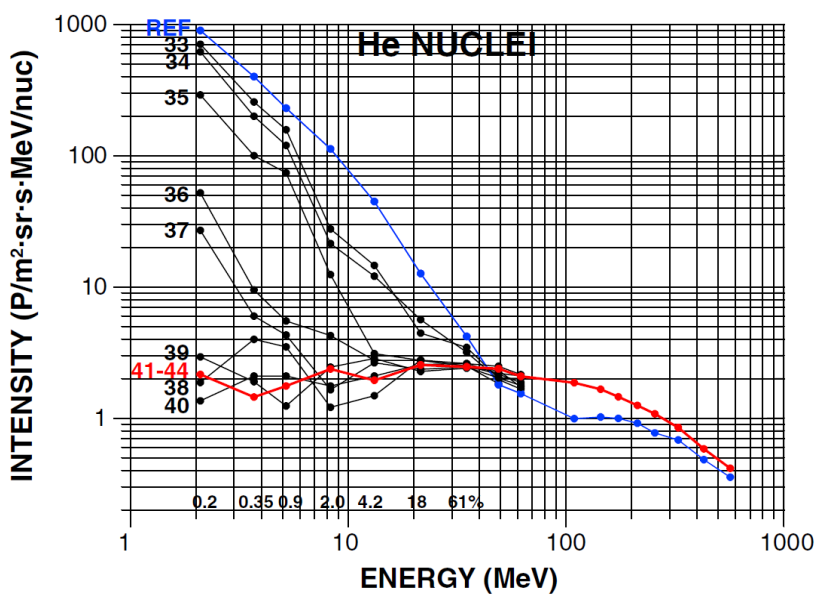}
			\caption{Weekly average helium spectra from 2 to 600 MeV/nuc observed at Voyager 1 during its crossing of what looks like the HP,
			from 2012.61 to the end of data (each week equals an outward movement of 0.07 AU). The time period 2011.8 to 2012.2 is used as a reference, 
			marked in blue, when Voyager 1 was still approaching the apparent HP. 
			Numbers to the left of each weekly average spectrum are the week numbers of the year 2012, with 41-44 in red representing a 4 week
			average as the intensity stabilized beyond the HP. Taken from Webber and McDonald [5] where a full discussion is given of these spectacular events.}
\label{fig:Figure3}
\end{figure*}

The question arises about what could cause such an increase in the electron spectrum in the mentioned energy range while a steady decrease is expected because of increasing solar modulation with decreasing kinetic energy. 
Because no process in the heliosphere is able to accelerate CR particles up to such high energies, it must be assumed that the appearance of such an unusual “bump” in the spectrum 
could most likely be ascribed to a galactic propagation and or acceleration process or perhaps to the presence of local source effects in the interstellar medium. 
Further investigation from an astrophysics point of view, and observational confirmation of this particular feature, are required. 

The power law at low energies was not explicitly predicted before
by any galactic propagation model. It seems likely that it can be obtained by simply adjusting the propagation parameters, e.g. the rigidity dependence of the diffusion process. The clear break between 
the two power laws may also follow because of a change in the propagation processes which are different at higher energies so that the spectral index of a electron source function may be steepened whereas at lower energies
it obviously does not happen. Another question to mention is what is the lowest energy that galactic electrons can have when reaching the heliopause? And at what decreasing energy will the $E^{-(1.5)}$ power law break down?

\begin{figure*}[!t]
\centering
\includegraphics[width=0.7\textwidth]{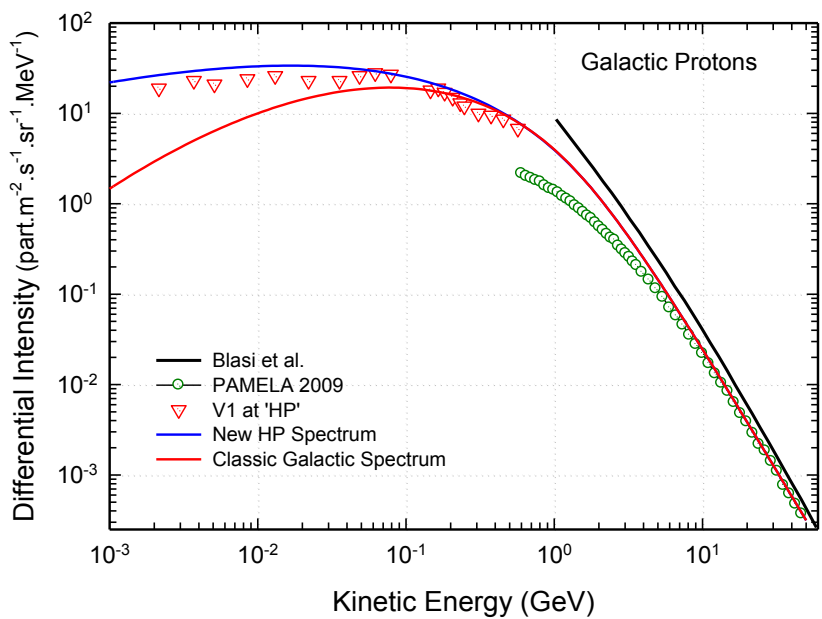}
			\caption{The newly calculated HPS for galactic protons (blue line) based on the proton spectrum observed below $\sim$500 MeV by Voyager 1 when it was exiting 
			the region that looks like the heliopause, assumingly entering the very local interstellar medium, and the PAMELA protons observations [6,7]
			at higher energies (green circles). Also shown as the red solid line is a typical GS based on a plain diffusion GALPROP model 
			as applied by Langner and Potgieter [28] to solar modulation. The black line is a LIS calculated by Blasi et al. [29].} 
\label{fig:Figure4}
\end{figure*}

\section{Galactic protons and helium}

The anomalous component of cosmic rays (ACRs) was discovered in the early 1970s. See the introductory review on this topic by Fichtner [25] and recent important developments by e.g. Giacalone et al. [26]. 
This component, with kinetic energy $E$ between $\sim$10 to 100~MeV/nuc, does not display the same spectral behaviour as galactic CRs but increases significantly with 
decreasing energy. Galactic CRs have harder spectra than ACRs. Their composition consists of hydrogen, helium, nitrogen, oxygen, neon and argon 
and is primarily singly ionized. They originate as interstellar neutrals that become ionized when flowing towards the Sun 
and then, as so-called pick-up ions, become accelerated in the heliosheath. To become ACRs, these pick-up ions 
must be accelerated by four orders of magnitude. They are subjected to solar modulation and depict mostly, but not always, the same modulation features 
than CRs upstream of the TS. Only the ACRs with the highest rigidity (oxygen) can reach Earth. 

The principal acceleration mechanism was considered to be diffusive shock acceleration, a topic of considerable debate since Voyager~1 crossed the TS [26]. 
At the location of the TS there was no direct evidence of the effective local acceleration of ACR protons but particles with lower energies were effectively accelerated 
and have since become known as termination shock particles (TSP). The higher energy ACRs thus seem disappointingly unaffected by the TS 
but had increased gradually in intensity away from the TS. They clearly gain energy as they move  outward inside the inner heliosheath and seem to be trapped largely in this region. 
Several very sophisticated mechanisms have been proposed [e.g. 26,27] of how these particles may gain their energy beyond the TS and has 
become one of the most severely debated issues in this field of research, an ongoing process.

Important for this review is that when Voyager 1 approached the heliopause region, one of the most spectacular observations of its history began to occur.
The TSP and ACR intensity had dropped significantly as the spacecraft exited the inner heliosheath, clearly illustrating that these particles are in fact heliospherically produced cosmic rays.
This is shown in figure 3 for helium, as an example, taken from Webber and McDonald [5]. This happened to all the other ACR species. From a very LIS point of view, this disappearance is equally spectacular
because now, for the first time, it has become possible to establish what a HPS for these galactic cosmic rays (without being contaminated by TSPs and ACRS) are at energies below $\sim$~100~MeV/nuc, as is shown in figure 3.

\begin{figure*}[!t]
\centering
\includegraphics[width=0.7\textwidth]{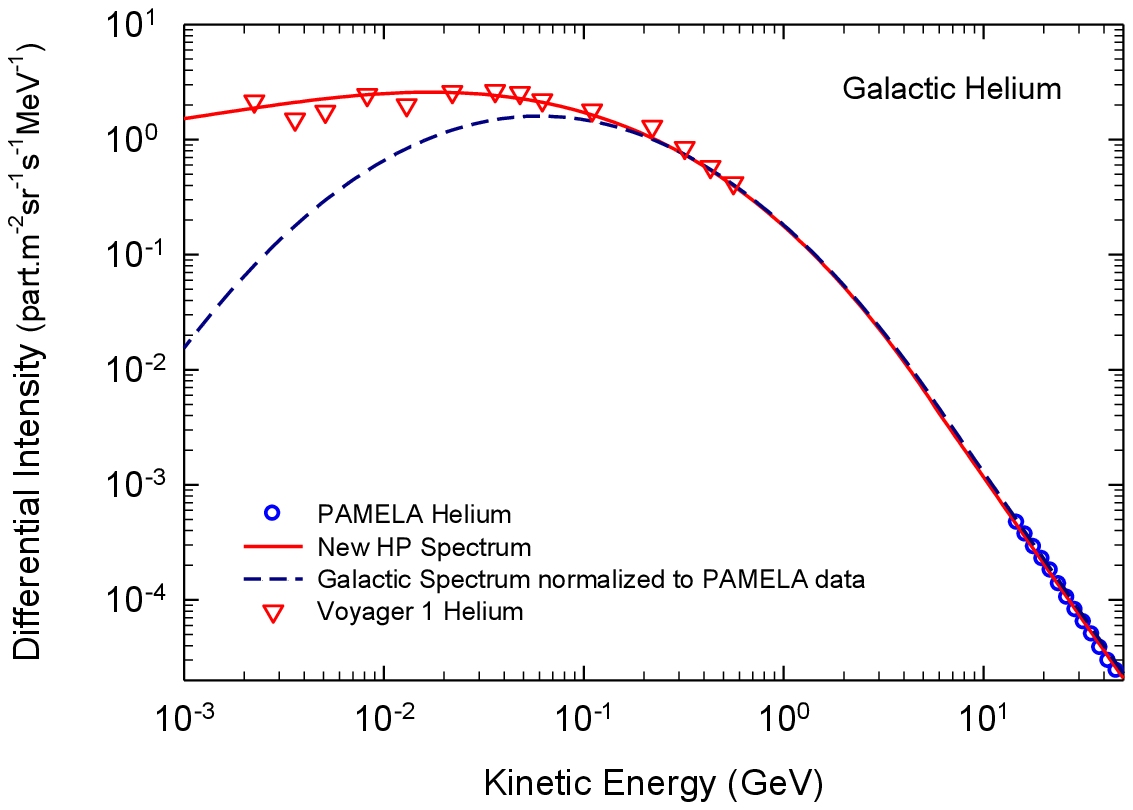}
			\caption{Similar to figure 4 but for galactic helium. Red solid line is the new HPS required to reproduce the Voyager 1 observations at low energies. At higher energies the HPS is 
			normalized to the PAMELA helium observations for 2009. This is compared to a typical GS based on a plain diffusion GALPROP model as applied by Langner and Potgieter [30]. As for protons,
			the difference below 200 MeV/nuc is significant. Clearly, this particular GALPROP modeling approach significantly underestimates the intensity of low energy protons and helium arriving at the heliosphere.
			}
\label{fig:Figure5}
\end{figure*}

In figure 4, the proton spectrum observed by Voyager 1 is shown, for the energy range as indicated, when it was exiting the region that looks like the heliopause and assumingly entering the very local
interstellar medium. This is compared to
a corresponding computed HPS including the PAMELA observations [6,7], shown in green. The red solid line is a typical GS as used many times before as an input spectrum for proton modulation studies. 
It is based on a plain diffusion GALPROP model as applied by Langner and Potgieter [28] to solar modulation modelling. The black line is a LIS computed by Blasi et al. [29], inferred from PAMELA data and Fermi-LAT gamma-ray 
observations of molecular clouds in the Gould belt. Evidently, the HPS is significantly higher at low energies than the classic type GS but differences disappear around 200 MeV. 

In figure 5, the observed HPS for helium is compared to a corresponding computed helium HPS and a GS computed with the diffusion version of GALPROP, and applied by 
Langner and Potgieter [30], but now normalising to the observed PAMELA spectra around 50 GeV/nuc. As for protons, this helium GS is significantly lower than the HPS for energies below 200 MeV/nuc.
Clearly, this specific galactic propagation model underestimates the intensity of these low energy protons and helium that arrive at the heliopause of the heliosphere.

Although only protons and helium spectra are shown here to illustrate the point that heliopause measurements now determine the very LIS at energies below a few hundred MeV,
a similar procedure can be followed for other cosmic ray species, such as carbon and oxygen.

\begin{figure*}[!t]
\centering
\includegraphics[width=0.65\textwidth]{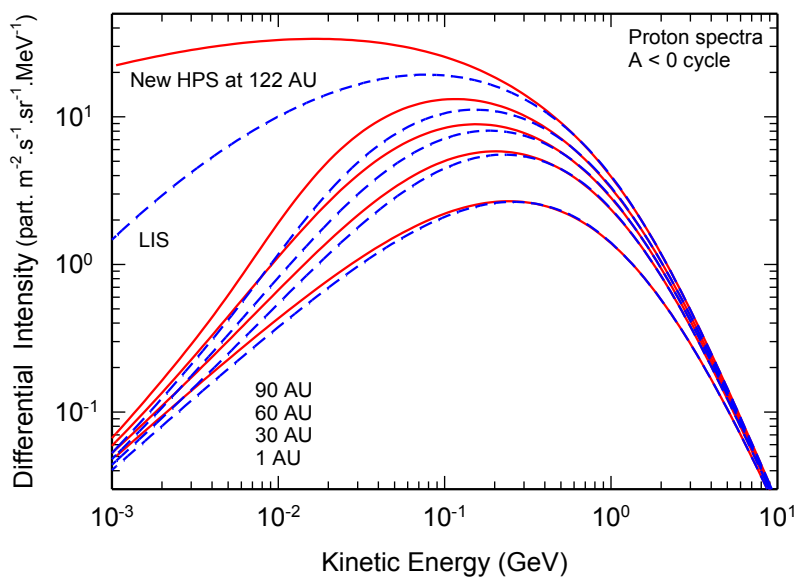}
		\caption{The new HPS for protons, specified at 122 AU, and based on Voyager 1 observations [13] at energies as indicated, is first used to compute modulated spectra 
		at 90 AU, 60 AU, 30 AU and at Earth, all indicated by red solid lines. This is then repeated for a GS as computed with a typical diffusion GALPROP model, all indicated by blue dashed lines.
		It shows how significantly very low energy protons are modulated, already between 122 AU and 90 AU but that at Earth the difference becomes small because the spectral shape is 
		determined by adiabatic energy losses.}
\label{fig:Figure6}
\end{figure*}

The question arises how the large difference between the HPS and a typical GS as shown in figure 4 for protons may affect the solar modulation from the outer to the inner heliosphere, up to Earth.
This is illustrated in figure 6, using the 3D modulation model of [15]. The new HPS, specified at 122 AU, is first used in the model to compute modulated spectra at 90 AU, 60 AU, 30 AU and at Earth, all indicated by red solid lines, 
then repeated for the GS, all indicated by blue dashed lines. It shows how significantly very low energy protons are modulated, already between 122 AU and 90 AU. 
Of course, 1 MeV protons arriving at Earth had significantly higher energy when they entered the heliosphere (see Strauss et al. for an illustration of this adiabatic energy loss effect [31]).  
For this reason, the computed spectra at Earth do not differ much. The relatively small difference around 200-300 MeV between the HPS and GS is mainly responsible for the difference at very low energies at Earth.
This again illustrates the point that observations at Earth below several GeV cannot accurately be used to determine a very LIS below a few GeV.

The computations shown in figure 6 allow to determine for the first time the total modulation that occurs in the heliosphere between the heliopause, assumed as modulation boundary,
and Earth. For galactic protons the reduction factor is 0.85 at 10 GeV, 0.73 at 5 GeV, 0.35 at 1 GeV, 0.09 at 100 MeV, and 0.014 at 10 MeV, and 0.0022 at 1 MeV. The latter corresponds to a global radial gradient
of  5 percent per AU in the equatorial nose direction of the heliosphere, whereas at 1 GeV it is 0.86 percent per AU.

\section{Conclusions}

An electron HPS, over an energy range from 5 MeV to 50 GeV, relevant to solar modulation, is presented that can be considered the lowest possible very LIS for electrons. 
This is done by using a comprehensive numerical model for solar modulation in comparison with Voyager 1 observations from 2010 and later, together with PAMELA 
spectra at Earth. Below $\sim$1.0 GeV, this HPS has a power law form with  $E^{-(1.50\pm0.15)}$. The uncertainty in the power index is mainly determined 
by the systematic uncertainties in the way the Voyager 1 electron spectra are calculated [13].

In order to reproduce the PAMELA electron spectrum for 2009, the HPS must however have a different power law form above $\sim$5 GeV, with $E^{-(3.15\pm0.05)}$. 
Combining the two power laws with a smooth transition from low to high energies produces the HPS shown in figure 1.

A second important result is that a break occurs in this HPS between $\sim$800 MeV and $\sim$2 GeV, consistent with the LIS results of Strong et al. [23],
who studied different GALPROP based models for the electron LIS, utilizing synchrotron radiation to constrain the low-energy interstellar electron spectrum.

The prediction is that the electron HP spectrum at 10 MeV, and at 122 AU from the Sun, has a value between 200 to 300 electrons m$^{-2}$s$^{-1}$sr$^{-1}$MeV$^{-1}$, which we consider 
as the lowest possible value of the LIS at this energy. The higher value seems consistent with the electron observations from Voyager 1 reported only very recently [13].
However, by studying the PAMELA electron spectrum from 2006 to the end of 2009, an unexpected ``bulge'' was found in the electron spectrum 
between $\sim$2 GeV and up to $\sim$20 GeV. This produces a spectral index of $-(3.3\pm 0.1)$ instead of  $-(3.15\pm 0.05)$ in the mentioned energy range. with the possibility
that two breaks are occurring in the electron spectrum. This new HPS is shown in figure 2. This feature cannot be caused by solar modulation or any other process inside the heliosphere.

In addition, a HPS for galactic protons and helium respectively is reported based on the recent Voyager 1 observations [13]. Comparing these HPS with previously calculated galactic spectra, 
shows significant difference below a few hundred MeV/nuc. It became evident that the typical GALPROP modeling approach significantly underestimates the intensity of low energy protons and helium arriving at the heliosphere.

For the first time, the total modulation that occurs in the heliosphere between the heliopause, assumed as modulation boundary, and Earth can be calculated accurately.

\vspace*{0.5cm}
\footnotesize{{\bf Acknowledgement:}{~The author thanks Mirko Boezio and his PAMELA colleagues for providing preliminary electron data. He also thanks his students, Etienne Vos and
Rendani Nndanganeni, for computational data. Partial financial support of the South African National Research Foundation (NRF) is acknowledged.}


\begin{thebibliography}{refs}

\bibitem{1} M.S. Potgieter, Living Rev. Solar Phys. 10 (2013) 3-66.
\bibitem{2} D.A. Gurnett, W.S Kurth, L.F. Burlaga and N.F. Ness, Sciencexpress 10.1126 (2013).
\bibitem{3} M.S. Potgieter and R.R. Nndanganeni, Astrophys. Space Sci. 345 (2013) 33-40.
\bibitem{4} G.S. Nkosi, M.S. Potgieter and W.R. Webber, Adv. Space Res. 48 (2011) 1480-1489.
\bibitem{5} W.R. Webber and F.B., McDonald, Geophys. Res. Lett. 40 (2013) 665-1668.
\bibitem{6} W. Menn, et al. Adv. Space Res. 51 (2013) 209-218.
\bibitem{7} O. Adriani, and PAMELA collaboration, Phys. Rev. Lett. 106 (2011) 201101:1-5.
\bibitem{8} V. Di Felice, Ph.D. thesis, University of Rome Tor Vergata, Italy (2010).
\bibitem{9} R. Munini, Master’s thesis, University of Trieste, Italy (2011).
\bibitem{10} E.E.Vos, Master’s thesis, North-West University, South Africa (2011).
\bibitem{11} M.S. Potgieter, E.E. Vos, R.R. Nndanganeni, M. Boezio and R. Munini, Proc. 33rd ICRC (2013).
\bibitem{12} M.S. Potgieter, R.R. Nndanganeni, E.E. Vos and M. Boezio, Proc. 33rd ICRC (2013).
\bibitem{13} E.C. Stone, A.C. Cummings, F.B. McDonald, B.C. Heikkila, N. Lal and W.R. Webber, Sciencexpress 10.1126 (2013).
\bibitem{14} S.M. Krimigis, R.B. Decker, E.C. Roelof, M.E. Hill, T.P. Armstrong, G. Gloeckler, D.C. Hamilton and L.J. Lanzerotti, Sciencexpress 10.1126 (2013).
\bibitem{15} M.S. Potgieter, E.E. Vos, M. Boezio, N. De Simone, V. Di Felice and V. Formato, Solar Phys. DOI 10.1007/s11207-013-0324-6 (2103)
\bibitem{16} M.S. Potgieter and R.R. Nndanganeni, Astropart. Phys. 48 (2013) 25-29.
\bibitem{17} S.E.S. Ferreira, K. Scherer, M.S. Potgieter, Adv. Space Res. 41 (2008) 351-360.
\bibitem{18} R.D. Strauss, M.S. Potgieter, S.E.S. Ferreira, H. Fichtner and K. Scherer, Astrophys. J. 765:L18 (2013) 1-6.
\bibitem{19} Xi Luo, M. Zhang, H.K. Rassoul, N.V. Pogorelov and J. Heerikhuisen, Astrophys. J. 764:85 (2013) 1-16.
\bibitem{20} E.N. Parker, Planet. Space Sci. 13 (1965) 9-49.
\bibitem{21} M.S. Potgieter and R.D. Strauss, Proc. 33rd ICRC (2013).
\bibitem{22} W.R. Webber and P.R. Higbie, J. Geophys. Res. 113 (2008) A11106:1-10.
\bibitem{23} A.W. Strong, E. Orlando and T.R. Jaffe, Astron. Astrophys. 534 (2011) A54:1-13.
\bibitem{24} K. Scherer, H. Fichtner, R.D. Strauss, S.E.S. Ferreira, M.S. Potgieter and H.-J. Fahr, Astrophys. J. 735:128 (2011) 1-5.
\bibitem{25} H. Fichtner, Space Sci. Rev. 95 (2001) 639-754.
\bibitem{26} J. Giacalone, J.F. Drake and J.R. Jokipii, Space Sci Rev 173 (2012) 283-307.
\bibitem{27} R.D. Strauss, M.S. Potgieter and S.E.S. Ferreira, Adv. Space Res. 48 (2011) 65-75.
\bibitem{28} U.W. Langner and M.S. Potgieter, Astrophys. J. 630 (2005) 1114-1124.
\bibitem{29} P. Blasi, E. Amato and P.D. Serpico, Physical Rev. Lett. 109 (2012) 061101:1-5.
\bibitem{30} U.W. Langner and M.S. Potgieter, Annales Geophys. 22 (2004) 3063-3072.
\bibitem{31} R.D. Strauss, M.S. Potgieter, A. Kopp and I. B\"{u}sching, J. Geophys. Res. 116 (2011) A12105:1-13.
\bibitem{32} U.W. Langner, O.C. de Jager and M.S. Potgieter, Adv. Space Res. 27 (2001), 517-522.

\end{thebibliography}
\end{document}